\documentclass[epj]{svjour}
\usepackage{graphics}
\begin{document}
\title{Eliminating cracking during drying}
\author{Qiu Jin\inst{1}, Peng Tan\inst{1}, Andrew B. Schofield\inst{2} \and Lei Xu\inst{1}
\thanks{e-mail:xulei@phy.cuhk.edu.hk}%
}                     
\institute{Department of Physics, The Chinese University of Hong
Kong, Hong Kong, China \and The School of Physics and Astronomy,
University of Edinburgh, Edinburgh, UK}
\date{Received: date / Revised version: date}
%
\abstract{ When colloidal suspensions dry, stresses build up and
cracks often occur - a phenomenon undesirable for important
industries such as paint and ceramics. We demonstrate an effective
method which can completely eliminate cracking during drying: by
adding emulsion droplets into colloidal suspensions, we can
systematically decrease the amount of cracking, and eliminate it
completely above a critical droplet concentration. Since the
emulsion droplets eventually also evaporate, our technique achieves
an effective function while making little changes to the component
of final product, and may therefore serve as a promising approach
for cracking elimination. Furthermore, adding droplets also varies
the speed of air invasion and provides a powerful method to adjust
drying rate. With the effective control over cracking and drying
rate, our study may find important applications in many drying and
cracking related industrial processes.
%
} 
\maketitle
\section{Introduction}
\label{sec:1} Drying of colloidal suspensions is closely related to
many important industrial processes, such as the drying of paints,
cosmetic products, and ceramic powders \cite{10}. During drying, the
``coffee-ring flow'' drives particles to fast-evaporating regions,
forming close packed networks \cite{1,deegan2}. Further evaporation
produces tiny liquid-air menisci between particles, resulting in
large capillary pressures that produce significant internal stresses
\cite{2,dufresne2,3}. Once the stress exceeds material strength,
cracking naturally occurs \cite{4}. Cracking is problematic for many
applications and thus its elimination is highly desirable. While the
occurrence, propagation, and patterns of cracking have been
extensively studied \cite{2,dufresne2,3,4,5,6,7,8,9}, the
fundamental mechanism of its formation remains open to debate. It
has been illustrated that the capillary pressure from air-liquid
interface is the source of internal stresses which cause cracking
\cite{2,dufresne2,3}. Equally important is the elastic modulus of
individual particles, which determines how particles respond to
these stresses \cite{Martinez,Singh,Singh2}. Therefore, manipulating
the internal structure, reducing stress and changing particle
modulus may achieve cracking reduction. For example, it was found
that combining multiple thin layers into a thick one can effectively
eliminate cracking \cite{Lee}. Another efficient approach is to add
soft particles into system, which reduces the effective modulus and
internal stress \cite{Singh2}. Despite this important finding,
however, adding soft particles can also bring disadvantages: these
particles may require complex synthesis and increase the cost; and
their addition may modify the component of final product. Therefore,
an effective approach with an easy-to-produce additive material
which makes little changes to the final product will be ideal.

In this study, we add ``liquid-particles'', i.e., emulsion droplets,
into the system, and achieve effective cracking elimination.
Interestingly, as the droplet concentration increases, a sharp
transition occurs in the cracking behavior, indicated by a dramatic
reduction of cracks. After this transition, the system becomes crack
free during drying, due to the decrease of the elastic modulus
($G'$); and the increasing importance of the loss modulus ($G''$).
Moreover, since the emulsion droplets eventually also evaporate, our
technique makes little modification to the component of final
product, and therefore suggests liquid droplets as a promising
additive material for cracking elimination. Using confocal
microscopy, we also find that droplets can slow down air invasion
and provide an effective control on the drying rate. With these
practical effects, our study may find useful applications in drying
and cracking related industrial processes.
\section{Experiment}
\label{sec:2}
\begin{figure}
\resizebox{0.45\textwidth}{!}{%
  \includegraphics{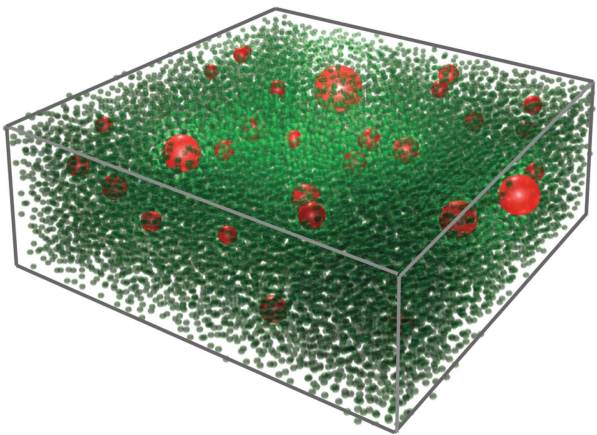}
} \caption{A schematic illustration of our complex fluid system,
courtesy of Peter Lu. The green monodispersed spheres indicate PMMA
colloids, and the red polydispersed spheres indicate emulsion
droplets. During drying, they form jammed structures as
illustrated.}
\label{fig:1}
\end{figure}
We suspend sterically-stabilized colloidal spheres of
polymethylmethacrylate (PMMA) with diameter $d = 300 nm $ in
decahydronaphthalene (DHN). We separately suspend a 1:1 (by volume)
mixture of water and glycerol, stabilized by PGPR-90 surfactant, in
DHN. We use a homogenizer to create a polydisperse emulsion, with
droplets ranging from hundreds of nanometers to a few microns. We
combine the suspensions of particles and of droplets at various
ratios to create our particle-droplet mixtures. We select these
particular components so that the refractive indices of the
particles, droplets and background solvent are all sufficiently
close that we can image the entire bulk with confocal microscopy. To
distinguish the particles from the droplets, we use different dyes
for each: particles are dyed with nitrobenzoxadiazole (NBD) and
appear green; droplets are dyed with rhodamine-B and appear red (see
Fig.1). We deposit $1.5\mu l$ of samples on a clean glass coverslip,
and image from below with confocal microscopy (Leica SP5). Drying
proceeds in two stages \cite{2}: in stage one, as the solvent
continues to evaporate, particles and droplets condense into a
jammed structure; during stage two, a drying air front invades the
jammed structure and displaces the solvent, which ultimately
evaporates completely. Cracking typically occurs at the beginning of
the second stage, during which internal stresses build up
dramatically.

\section{Results and discussion}
\label{sec:2}
\begin{figure}
\resizebox{0.48\textwidth}{!}{%
  \includegraphics{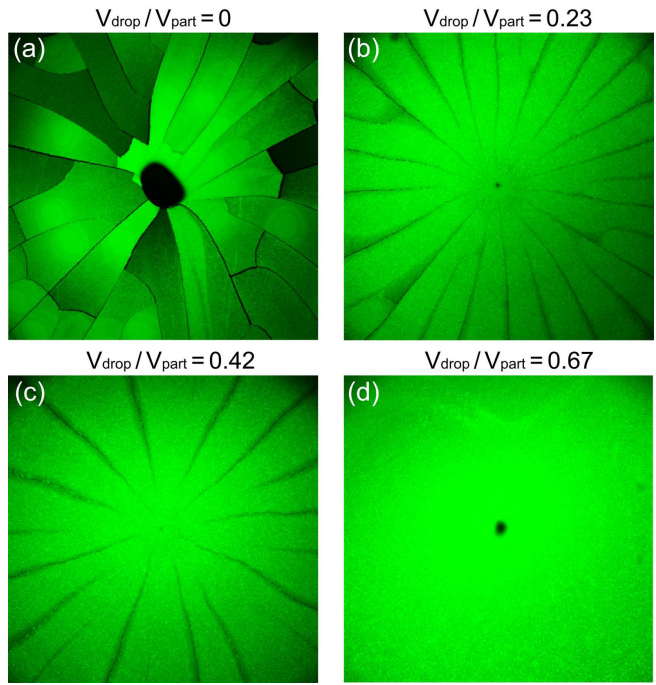}
} \caption{Reducing cracks by adding emulsion droplets. The total
dimensions of each image are $1.55\times1.55 mm^2$. (a) for pure
particle suspension with no droplets, there are a large number of
cracks distributing irregularly throughout the sample. (b) with
small amount of droplets ($V_{drop}/V_{part}=0.23$), plenty of
cracks still arise, but distribute more regularly along radial
directions. (c) as more emulsion droplets are added
($V_{drop}/V_{part}=0.42$), the amount of cracks decreases
significantly. (d) with enough droplets ($V_{drop}/V_{part}=0.67$),
cracks disappear completely.} \label{fig:2}
\end{figure}
We demonstrate the cracking results of several samples, with
increasing droplet concentrations, in Fig.2. Due to the continuous
evaporation of solvent, the volume fractions of droplets and
particles keep increasing and thus can not be taken as meaningful
parameters. However, the volume ratio between droplets and
particles, $V_{drop}/V_{part}$, remains a constant during
measurement \cite{footnote}. We therefore specify the samples with
$V_{drop}/V_{part}$. Clearly, the cracking behavior varies
systematically with $V_{drop}/V_{part}$. For the pure colloidal
suspension, a large number of cracks distribute irregularly
throughout the sample (Fig.2a, $V_{drop}/V_{part}$ = $0$). When
small amount of droplets are added, plenty of cracks still arise,
but distribute more regularly along radial directions (Fig.2b,
$V_{drop}/V_{part}=0.23$). As more and more droplets are included,
the amount of cracks decreases dramatically (Fig.2c,
$V_{drop}/V_{part}=0.42$), until they disappear completely (Fig.2d,
$V_{drop}/V_{part}=0.67$). Apparently, adding droplets effectively
reduces the cracking formation, regularizes the cracking
distribution, and eventually leads to a crack free system.

\begin{figure}
\resizebox{0.48\textwidth}{!}{%
  \includegraphics{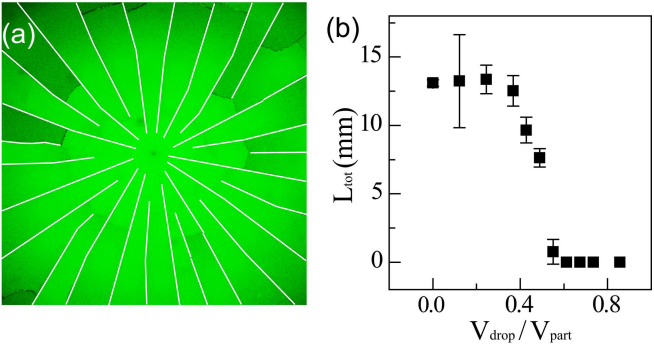}
} \caption{The total crack length, $L_{tot}$, versus
droplet-particle ratio, $V_{drop}/V_{part}$. (a) measuring $L_{tot}$
in a typical field of view ($1.55\times1.55 mm^2.$). We sum up the
total length of cracks within the field of view, as illustrated by
the white lines. To make consistent comparisons, we only count
cracks along radial directions. (b) $L_{tot}$ vs.
$V_{drop}/V_{part}$. Each data point is determined by the average of
three samples at the same $V_{drop}/V_{part}$ ratio, with error bars
calculated from the standard deviation. Apparently, for samples with
zero or small $V_{drop}/V_{part}$ ratio, $L_{tot}$ remains largely
unvaried at a high level. However, above a critical ratio around
0.4, $L_{tot}$ quickly drops to zero, and remains at zero
afterwards. The data indicate that the cracking reduction does not
occur gradually; instead it undergoes a critical-like transition
near a threshold droplet-particle ratio.}
\label{fig:3}
\end{figure}

To quantify the degree of cracking, we measure the total length of
all cracks within the field of view, as illustrated by the white
lines in Fig.3a. We compare this total length, $L_{tot}$, for
samples with different $V_{drop}/V_{part}$ in Fig.3b. To make
consistent comparisons, we always put our field of view at the
center of each sample, and count the major cracks along radial
directions only. Each $L_{tot}$ is determined by the average of
three experiments at the same $V_{drop}/V_{part}$, with error bars
calculated from the standard deviation. Clearly, for samples with
zero or small droplet amount, $L_{tot}$ remains largely unvaried at
a high level. However, above a critical ratio around
$V_{drop}/V_{part}=0.4$, $L_{tot}$ rapidly drops to zero, and
remains at zero afterwards. These data indicate that the cracking
reduction does not occur gradually; instead it undergoes a
critical-like transition around a threshold $V_{drop}/V_{part}$
value.

How to understand this critical transition? We investigate the
samples' fundamental mechanical property -- the viscoelastic moduli
-- with respect to $V_{drop}/V_{part}$. Unlike pure solids or
liquids, which respond to external stress with either elastic
deformation or viscous dissipation; colloidal suspensions exhibit
both elastic (or solid-like) and viscous (or liquid-like) responses.
These responses are quantitatively described by the two viscoelastic
moduli: the storage modulus, $G'(\omega)$, and the loss modulus,
$G''(\omega)$. Their relative importance determines the amount of
energy stored versus dissipated. We directly measure $G'(\omega)$
and $G''(\omega)$ for samples with different $V_{drop}/V_{part}$.
Since in our experiment, cracks appear after the system being under
stress for about $20s$, we use $f=0.05s^{-1}$ as the frequency lower
bound. The upper bound is chosen as $f=30s^{-1}$, which is the
typical formation time of one individual crack.

\begin{figure}
\resizebox{0.45\textwidth}{!}{%
  \includegraphics{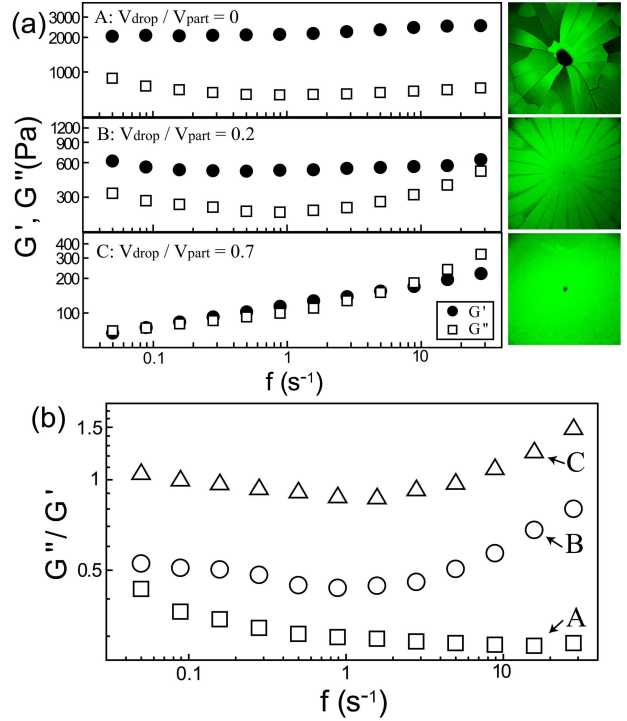}
}
\caption{Characterizing the storage modulus, $G'$, and loss
modulus, $G''$, at different droplet-particle ratios. (a) The
decrease of $G'$ and $G''$ with the increase of droplet
concentration. In addition, for sample A with no droplets, $G'$ is
significantly larger than $G''$, indicating that most energy can be
elastically stored. In sample B, a small amount of droplets are
added but at a concentration smaller than the critical transition.
$G'$ still dominates $G''$ but with a smaller difference. For sample
C with high droplet concentration, $G'$ and $G''$ have similar
magnitude and significant fraction of energy can dissipate through
$G''$. The images on the right illustrate the corresponding cracking
behaviors. (b) the ratio between the two moduli, $G''/G'$, for the
same data in (a). As more droplets are added, $G''/G'$ increases
from much smaller than one to close to one, demonstrating the
growing importance of $G''$} \label{fig:4}
\end{figure}

Three typical samples are measured: sample A without emulsion
droplets ($V_{drop}/V_{part}=0$), sample B with small amount of
droplets before the critical transition ($V_{drop}/V_{part}$ =
$0.2$), and sample C with large amount of droplets after the
transition ($V_{drop}/V_{part}=0.7$). All samples have similar
volume fractions around $60\pm3\%$. This volume fraction is
approximately the same as the jammed structure right before air
invades the system. Their viscoelastic moduli, $G'(\omega)$ and
$G''(\omega)$, are shown in Fig.4a. From sample A to C, $G'$ and
$G''$ decrease with $V_{drop}/V_{part}$. The decrease of $G'$ may
reduce the internal stress in the system and cause cracking
reduction, consistent with previous findings \cite{Singh2}.

Moreover, we also observe a systematic change in the relative
importance between $G'$ and $G''$: for sample A, $G'(\omega)$ is
significantly higher than $G''(\omega)$, indicating that under
external stresses most energy will be elastically stored within the
system. For sample B, $G'(\omega)$ still dominates $G''(\omega)$ but
with a smaller difference, suggesting the increasing importance of
$G''$. When enough droplets are added in sample C, $G'(\omega)$ and
$G''(\omega)$ become to have similar magnitude and significant
fraction of energy can dissipate through $G''$. We therefore plot
the ratio, $G''(\omega)/G'(\omega)$, at different
$V_{drop}/V_{part}$ in Fig.4b: as the droplet amount increases,
$G''(\omega)/G'(\omega)$ grows from significantly smaller than one
(sample A) to close to unity (sample C), indicating that more
fraction of energy can dissipate through $G''$ while less can be
stored through $G'$. Therefore, our experiment indicates that the
cracking reduction is related to both the decrease of $G'$ and the
increasing importance of $G''$, suggesting the significant roles of
both factors.

\begin{figure}
\resizebox{0.48\textwidth}{!}{%
  \includegraphics{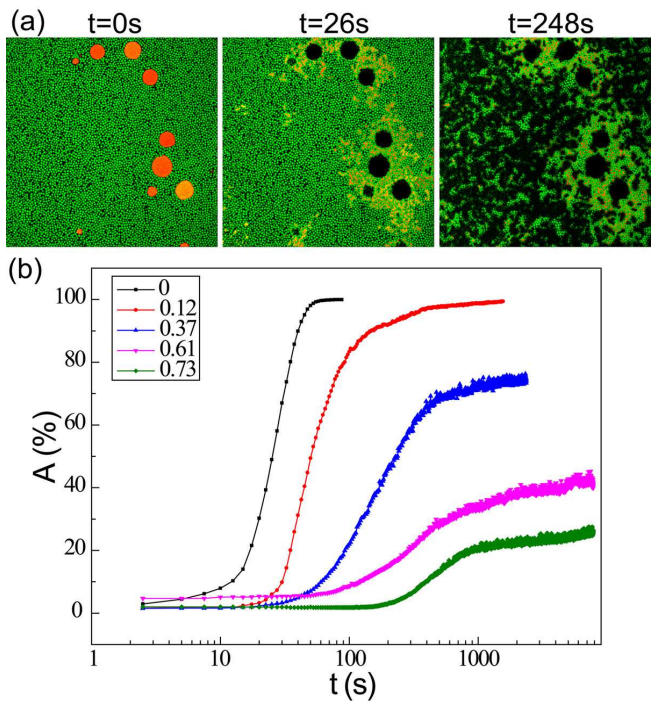}
} \caption{(Color online) Varying the air invasion speed by adding
emulsion droplets. (a) air invasion visualized by confocal
microscopy. To make single-particle level measurement, we use large
particles ($d=3\mu m$) and droplets ($d'=10\sim100\mu m$). When air
enters system and replaces solvent, the index match is destroyed and
the invaded region turns into black. During invasion, the droplet
contents are pushed into surrounding pores, as shown by the second
image. More interestingly, these contents can `protect' the nearby
regions from being invaded, as demonstrated in the third image. (b)
the fraction of invaded area, $A(t)$, for samples at different
droplet-particle ratios. For the sample with no droplets, it only
takes about $50s$ to reach the stable $A(t)$ value at $100\%$; while
for the one with $V_{drop}/V_{part}=0.73$, it takes about $1000s$ to
reach the stable $A(t)$ at $23\%$ (The decrease of stable $A(t)$
values is due to the increase of the almost low-volatile droplet
contents). This significant variation in the speed of air invasion
may provide a powerful control on the drying rate for practical
applications.} \label{fig:5}
\end{figure}

More interestingly, adding droplets into system has another effect:
it can slow down the invasion speed of air and achieve powerful
control on drying rate. The air invasion can be directly measured by
confocal microscopy shown in Fig.5a: as air invades the system, more
and more regions turn into black. This is due to the replacement of
solvent by air, which destroys the refractive index match and makes
the region appear black. To quantitatively describe the invasion
speed, we continuously scan a fixed field of view ($0.91\times0.91
mm^2$) at the frame rate of $1.3$ frame/s, and measure the
percentage of invaded area as a function of time, $A(t)$. We compare
$A(t)$ for samples with different $V_{drop}/V_{part}$ in Fig.5b.
Here the time zero is defined as the first frame at which air enters
the field of view. Clearly, air invasion slows down dramatically
with the increase of $V_{drop}/V_{part}$. For the sample without
droplets, it only takes about $50s$ for $A(t)$ to reach the stable
value of $100\%$; while for $V_{drop}/V_{part}=0.73$, it takes more
than $1000s$ to reach the stable value of $23\%$. To clarify whether
the reduction of cracking is related to the slow down of air
invasion, we compare the drying of the same samples both in open air
and inside a chamber. For the latter situation, the air invasion
speed decreases by more than eight times; while similar amount of
cracks still arises. Therefore, in our system evaporation rate of
solvent does not have a significant influence on cracking
elimination, and the reduction is mainly due to the effect from
droplets.

Why droplets can slow down the air invasion? To understand it, we
investigate the invasion process at single-particle level. We
prepare a droplet-particle mixture with particle diameter $d=3\mu m$
and droplet diameter $d'=10\sim100\mu m$, and inspect the detailed
invasion process in Fig.5a. Once touched by air, the droplets
collapse and their contents are pushed into surrounding pores
(Fig.5a middle image), due to the large pressure imbalance
previously discovered \cite{11}. Interestingly, as air continues to
enter the system, these contents can block the pathways of air and
prevent nearby regions from being invaded. This phenomenon is
clearly demonstrated in Fig.5a: the regions far away from the
droplets get quickly invaded by air and turn into black; while the
regions near the droplets remain bright and not invaded for a long
period of time. Therefore, when enough droplets are added,
significant fraction of pore space can be `protected' by droplet
contents which can slow down the air invasion considerably. We
emphasize that the droplet contents eventually also evaporate,
leaving only a little surfactant. Therefore adding droplets can
achieve both cracking reduction and drying rate variation without
significantly altering the final product.

The observation that the low-volatile droplet contents in pore space
can dramatically slow down the air invasion naturally raises another
possible explanation for our crack elimination: if the pore space is
totally occupied by these contents, air can hardly enter and crack
thus disappears. However, we point out two issues in this
explanation: (1) the low-volatile contents eventually also evaporate
and air sooner or later will enter the system, thus these contents
should just delay instead of eliminate cracking; and (2) we perform
new experiments with pure water as droplet contents and find similar
crack elimination behaviors. Since water is even more volatile than
the bulk solvent and thus does not slow down air invasion, this
finding shows that the low-volatility of droplet contents is
probably not the reason. Therefore these results suggest that the
the change of moduli provides a more reasonable explanation for our
cracking elimination.

\section{Conclusion}
In this study, we experimentally illustrate an effective approach
for cracking elimination during drying: by adding emulsion droplets,
the amount of cracks decreases systematically, until they are
completely eliminated. Moreover, since the droplets eventually also
evaporate, our approach makes little modification to the component
of final product, and thus suggests liquid droplets as one
potentially powerful candidate for cracking elimination. More
explorations along this direction may bring effective and cheap
cracking elimination methods. In addition, droplets can also slow
down the air invasion and provide an effective control on the drying
rate, which has practical importance in paint and cosmetics
industries.

\begin{acknowledgement}
This project is supported by the Hong Kong GRF Grant (Project No.
CUHK404211).
\end{acknowledgement}
%

\begin{thebibliography}{}
\bibitem{footnote} The evaporation rate of the droplets is much slower than the solvent
therefore the droplet volume is approximately a constant during our
measurement time.
\end{thebibliography}
%

\end{document}